\documentclass[ALICE,manyauthors]{cernphprep}
\usepackage[comma,square,numbers,sort&compress]{natbib}

\usepackage{lineno}
\usepackage{xspace}
\usepackage{hyperref}
\usepackage{color}
\usepackage{amssymb}
\usepackage{gensymb}
\usepackage[T1]{fontenc}
\usepackage{multirow}
\usepackage{makecell}
\usepackage{tabularx}
\usepackage{xcolor}
\usepackage{xspace}

\begin{document}
%

\newcommand{\pp}           {pp\xspace}
\newcommand{\ppbar}        {\mbox{$\mathrm {p\overline{p}}$}\xspace}
\newcommand{\XeXe}         {\mbox{Xe--Xe}\xspace}
\newcommand{\PbPb}         {\mbox{Pb--Pb}\xspace}
\newcommand{\pA}           {\mbox{pA}\xspace}
\newcommand{\pPb}          {\mbox{p--Pb}\xspace}
\newcommand{\AuAu}         {\mbox{Au--Au}\xspace}
\newcommand{\dAu}          {\mbox{d--Au}\xspace}

\newcommand{\s}            {\ensuremath{\sqrt{s}}\xspace}
\newcommand{\snn}          {\ensuremath{\sqrt{s_{\mathrm{NN}}}}\xspace}
\newcommand{\pt}           {\ensuremath{p_{\rm T}}\xspace}
\newcommand{\meanpt}       {$\langle p_{\mathrm{T}}\rangle$\xspace}
\newcommand{\ycms}         {\ensuremath{y_{\rm CMS}}\xspace}
\newcommand{\ylab}         {\ensuremath{y_{\rm lab}}\xspace}
\newcommand{\etarange}[1]  {\mbox{$\left | \eta \right |~<~#1$}}
\newcommand{\yrange}[1]    {\mbox{$\left | y \right |~<~#1$}}
\newcommand{\dndy}         {\ensuremath{\mathrm{d}N_\mathrm{ch}/\mathrm{d}y}\xspace}
\newcommand{\dndeta}       {\ensuremath{\mathrm{d}N_\mathrm{ch}/\mathrm{d}\eta}\xspace}
\newcommand{\avdndeta}     {\ensuremath{\langle\dndeta\rangle}\xspace}
\newcommand{\dNdy}         {\ensuremath{\mathrm{d}N_\mathrm{ch}/\mathrm{d}y}\xspace}
\newcommand{\Npart}        {\ensuremath{N_\mathrm{part}}\xspace}
\newcommand{\Ncoll}        {\ensuremath{N_\mathrm{coll}}\xspace}
\newcommand{\dEdx}         {\ensuremath{\textrm{d}E/\textrm{d}x}\xspace}
\newcommand{\RpPb}         {\ensuremath{R_{\rm pPb}}\xspace}

\newcommand{\nineH}        {$\sqrt{s}~=~0.9$~Te\kern-.1emV\xspace}
\newcommand{\seven}        {$\sqrt{s}~=~7$~Te\kern-.1emV\xspace}
\newcommand{\twoH}         {$\sqrt{s}~=~0.2$~Te\kern-.1emV\xspace}
\newcommand{\twosevensix}  {$\sqrt{s}~=~2.76$~Te\kern-.1emV\xspace}
\newcommand{\five}         {$\sqrt{s}~=~5.02$~Te\kern-.1emV\xspace}
\newcommand{\twosevensixnn}{$\sqrt{s_{\mathrm{NN}}}~=~2.76$~Te\kern-.1emV\xspace}
\newcommand{\fivenn}       {$\sqrt{s_{\mathrm{NN}}}~=~5.02$~Te\kern-.1emV\xspace}
\newcommand{\LT}           {L{\'e}vy-Tsallis\xspace}
\newcommand{\GeVc}         {Ge\kern-.1emV/$c$\xspace}
\newcommand{\MeVc}         {Me\kern-.1emV/$c$\xspace}
\newcommand{\TeV}          {Te\kern-.1emV\xspace}
\newcommand{\GeV}          {Ge\kern-.1emV\xspace}
\newcommand{\MeV}          {Me\kern-.1emV\xspace}
\newcommand{\GeVmass}      {Ge\kern-.2emV/$c^2$\xspace}
\newcommand{\MeVmass}      {Me\kern-.2emV/$c^2$\xspace}
\newcommand{\lumi}         {\ensuremath{\mathcal{L}}\xspace}

\newcommand{\ITS}          {\rm{ITS}\xspace}
\newcommand{\TOF}          {\rm{TOF}\xspace}
\newcommand{\ZDC}          {\rm{ZDC}\xspace}
\newcommand{\ZDCs}         {\rm{ZDCs}\xspace}
\newcommand{\ZNA}          {\rm{ZNA}\xspace}
\newcommand{\ZNC}          {\rm{ZNC}\xspace}
\newcommand{\SPD}          {\rm{SPD}\xspace}
\newcommand{\SDD}          {\rm{SDD}\xspace}
\newcommand{\SSD}          {\rm{SSD}\xspace}
\newcommand{\TPC}          {\rm{TPC}\xspace}
\newcommand{\TRD}          {\rm{TRD}\xspace}
\newcommand{\VZERO}        {\rm{V0}\xspace}
\newcommand{\VZEROA}       {\rm{V0A}\xspace}
\newcommand{\VZEROC}       {\rm{V0C}\xspace}
\newcommand{\Vdecay} 	   {\ensuremath{V^{0}}\xspace}

\newcommand{\ee}           {\ensuremath{e^{+}e^{-}}} 
\newcommand{\pip}          {\ensuremath{\pi^{+}}\xspace}
\newcommand{\pim}          {\ensuremath{\pi^{-}}\xspace}
\newcommand{\kap}          {\ensuremath{\rm{K}^{+}}\xspace}
\newcommand{\kam}          {\ensuremath{\rm{K}^{-}}\xspace}
\newcommand{\pbar}         {\ensuremath{\rm\overline{p}}\xspace}
\newcommand{\kzero}        {\ensuremath{{\rm K}^{0}_{\rm{S}}}\xspace}
\newcommand{\lmb}          {\ensuremath{\Lambda}\xspace}
\newcommand{\almb}         {\ensuremath{\overline{\Lambda}}\xspace}
\newcommand{\Om}           {\ensuremath{\Omega^-}\xspace}
\newcommand{\Mo}           {\ensuremath{\overline{\Omega}^+}\xspace}
\newcommand{\X}            {\ensuremath{\Xi^-}\xspace}
\newcommand{\Ix}           {\ensuremath{\overline{\Xi}^+}\xspace}
\newcommand{\Xis}          {\ensuremath{\Xi^{\pm}}\xspace}
\newcommand{\Oms}          {\ensuremath{\Omega^{\pm}}\xspace}
\newcommand{\degre}{\ensuremath{^{\rm o}}\xspace}
\newcommand{\XicZero}{\ensuremath{\Xi_{\rm c}^0}\xspace}
\newcommand{\XicPlus}{\ensuremath{\Xi_{\rm c}^+}\xspace}

\newcommand{\epem}         {\mbox{$\mathrm {e^{+}e^{-}}$}\xspace}
\newcommand{\ep}           {\mbox{$\mathrm{e}^{-}\mathrm{p}$}\xspace}
\newcommand{\thirteen}     {$\sqrt{s}=13$~Te\kern-.1emV\xspace}
\newcommand{\thirteensix}  {$\sqrt{s}~=~13.6$~Te\kern-.1emV\xspace}
\newcommand{\VZEROM}       {\rm{V0M}\xspace}

\newcommand{\ptrange}[2]   {\mbox{$#1 < {p_{\rm T}} < #2$}}

\newcommand{\pizero}       {\ensuremath{\pi^{0}}\xspace}
\newcommand{\Lambdab}      {\ensuremath{\Lambda_{\rm b}^{0}}\xspace}
\newcommand{\Lambdac}      {\ensuremath{\Lambda_{\rm c}^{+}}\xspace}
\newcommand{\Xib}          {\ensuremath{\Xi_{\rm b}}\xspace}
\newcommand{\Xic}          {\ensuremath{\Xi_{\rm c}}\xspace}
\newcommand{\Xicplus}      {\ensuremath{\Xi_{\rm c}^{+}}\xspace}
\newcommand{\Xiczero}      {\ensuremath{\Xi_{\rm c}^{0}}\xspace}
\newcommand{\Xiczeroplus}  {\ensuremath{\Xi_{\rm c}^{0,+}}\xspace}
\newcommand{\Sigmac}       {\ensuremath{\Sigma_{\rm c}^{0,++}\mathrm{(2455)}}\xspace}
\newcommand{\Dzero}        {\ensuremath{\rm D^{0}}\xspace}
\newcommand{\Ds}           {\ensuremath{\rm D^{+}_{s}}\xspace}
\newcommand{\Dplus}        {\ensuremath{\rm D^{+}}\xspace}
\newcommand{\Dstar}        {\ensuremath{\rm D^{*+}}\xspace}
\newcommand{\Omegac}       {\ensuremath{\Omega_{\rm c}^{0}}\xspace}
\newcommand{\nue}          {\ensuremath{\nu_{e}}\xspace}
\newcommand{\eplus}        {\ensuremath{\rm e^{+}}\xspace}

\newcommand{\acceff}       {$(\rm{Acc} \times \epsilon)$\xspace}
\newcommand{\XiePair}      {\ensuremath{\rm e\Xi}\xspace}
\newcommand{\ntrk}         {$N_{tracklet}$\xspace}

\newcommand{\slfrac}[2]{\left.#1\right/#2}
\newcommand{\XicZeroToPiXi}{$\rm \Xi_{c}^{0} \rightarrow \Xi^{-} \pi^{+}$\xspace}
\newcommand{\XicZeroToXiEleNu}{$\Xiczero \rightarrow \Xi^- {\rm e}^+ \rm \nu_{\rm e}$\xspace}
\newcommand{\XicPlusToXiPiPi}{$\rm \Xi_{c}^{+} \rightarrow \Xi^{-} \pi^{+} \pi^{+}$\xspace}
\newcommand{\pvom}{$\it{p}_{\mathrm{V0M}}$\xspace}

\begin{titlepage}
\PHyear{2021}       
\PHnumber{084}      
\PHdate{10 May}  

\title{Measurement of the cross sections of $\Xi^{0}_{\rm c}$ and $\Xi^+_{\rm c}$ baryons and of the branching-fraction ratio ${\rm BR}(\Xi^0_{\rm c} \rightarrow \Xi^{-}{\rm e}^{+}\nu_{\rm e})/ {\rm BR}(\Xi^0_{\rm c} \rightarrow \Xi^{-}{\pi}^{+})$ in pp collisions at $\sqrt{s}=$~13~TeV}
\ShortTitle{$\Xi^0_{\rm c}$ and $\Xi^+_{\rm c}$ production in pp collisions at $\sqrt{s}=$~13~TeV}   

\Collaboration{ALICE Collaboration\thanks{See Appendix~\ref{app:collab} for the list of collaboration members}}
\ShortAuthor{ALICE Collaboration} 

\begin{abstract}
The $\pt$-differential cross sections of prompt charm-strange baryons $\Xi^0_{\rm c}$
and $\Xi^+_{\rm c}$ were measured at midrapidity ($|y| < 0.5$) in proton--proton (pp) collisions at a centre-of-mass energy $\sqrt{s}=$~13~TeV with the ALICE detector at the LHC. The $\Xi^0_{\rm c}$ baryon was reconstructed via both the semileptonic decay ($\Xi^{-}{\rm e^{+}}\nu_{\rm e}$) and the hadronic decay ($\Xi^{-}{\rm \pi^{+}}$) channels. The $\Xi^+_{\rm c}$ baryon was reconstructed via the hadronic decay ($\Xi^{-}\pi^{+}\pi^{+}$) channel.
The branching-fraction ratio $\rm {\rm BR}(\Xi_c^0\rightarrow~ \Xi^-e^+\nu_e)/\rm {\rm BR}(\Xi_c^0\rightarrow \Xi^{-}\pi^+)=$ 0.95 $\pm$ 0.15 (stat) $\pm$ 0.16 (syst) was consistent with the Belle's result within 1$\sigma$.
The transverse momentum ($\pt$) dependence of the $\Xi^0_{\rm c}$- and $\Xi^+_{\rm c}$-baryon production relative to the ${\rm D^0}$-meson and to the $\Sigma^{0,+,++}_{\rm c}$- and $\Lambda^+_{\rm c}$-baryon production are reported. The baryon-to-meson ratio increases towards low \pt up to a value of approximately 0.3.
The measurements are compared with various models that take different hadronisation mechanisms into consideration. The results provide stringent constraints to these theoretical calculations and additional evidence that different processes are involved in charm hadronisation in electron--positron ($\rm e^+e^-$) and hadronic collisions.

\end{abstract}
\end{titlepage}

\setcounter{page}{2} 

Measurements of heavy-flavour hadron production in high-energy proton--proton (pp) collisions provide important tests of quantum chromodynamics (QCD). The cross sections of heavy-flavour hadrons are usually computed using the factorisation approach as a convolution of three factors~\cite{Collins:1985gm}: i) the parton distribution functions of the incoming protons, ii) the hard-scattering cross section at partonic level, and iii) the fragmentation function of heavy quarks into a given heavy-flavour hadron.
The D- and B-meson cross sections in pp collisions at several centre-of-mass energies at the LHC~\cite{Acharya:2019mgn,ALICE:2021mgk,Sirunyan:2017xss,Khachatryan:2011mk,Chatrchyan:2011pw,Chatrchyan:2011vh} are described within uncertainties by perturbative QCD calculations~\cite{Kramer:2017gct,Helenius:2018uul,Cacciari:1998it,Cacciari:2012ny,Kniehl:2020szu}, which use fragmentation functions tuned on $\rm{e^+e^-}$ data, over a wide range of transverse momentum (\pt).
Measurements of $\rm \Lambda^{+}_c$-baryon production at midrapidity in pp collisions at the centre-of-mass energy $\sqrt{s}=$~5.02 and 7~TeV were reported by the ALICE and CMS Collaborations in Refs.~\cite{Sirunyan:2019fnc,Acharya:2017kfy,Acharya:2020uqi}. The measured $\rm \Lambda^{+}_c/{\rm D^0}$ ratio is higher than previous measurements in $\rm{e^+e^-}$~\cite{Albrecht:1988an,Avery:1990bc,Gladilin:2014tba} and ${\rm e^-p}$~\cite{Chekanov:2005mm,Abramowicz:2013eja} collisions. 
A similar observation was drawn from the measurement of the inclusive $\Xi^0_{\rm c}$-baryon production at midrapidity in pp collisions at $\sqrt{s} = 7$~TeV~\cite{Acharya:2017lwf}.

PYTHIA~8.2 tunes including string formation beyond the leading-colour approximation~\cite{Christiansen:2015yqa} and a statistical hadronisation model (SHM)~\cite{He:2019tik} including a set of higher-mass charm-baryon states as prescribed by the relativistic quark model (RQM) and from lattice QCD~\cite{Ebert:2011kk,Briceno:2012wt} qualitatively describe the measured $\rm \Sigma^{0,+,++}_c/{\rm D^0}$ and $\rm \Lambda^{+}_c/{\rm D^0}$ cross section ratios~\cite{SigmacLambdac,Acharya:2020uqi}, but underestimate the $\Xi^0_{\rm c}/{\rm D^0}$ ratio~\cite{Acharya:2017lwf}. 
The observed enhancement of the charm-baryon production can also be explained by model calculations considering hadronisation of charm quarks via coalescence in addition to the fragmentation  in pp collisions~\cite{Song:2018tpv,Minissale:2020bif}.
The increased yield of charm baryons makes it mandatory to include their contribution for an accurate measurement of the ${\rm c\overline{c}}$ production cross section in pp collisions at the LHC~\cite{Acharya:2017jgo}. 

In this Letter, the measurements of the cross sections of the prompt (i.e., produced directly in the hadronisation of charm quarks and in the decays of directly produced excited charm states) charm-strange baryons $\Xi^0_{\rm c}$ and $\Xi^+_{\rm c}$ at midrapidity ($|y|<0.5$) in pp collisions at $\sqrt{s} = 13$~TeV are reported. 
The $\Xi^0_{\rm c}$ baryon was reconstructed via the decay channels ${\Xi^{-}\rm e^{+}}\nu_{\rm e}$, BR $=$ ($1.8\pm 1.2$)\% and $\Xi^{-}{\rm \pi^{+}}$, BR $=$ (1.43 $\pm$ 0.32)\%~\cite{Zyla:2020zbs} together with their charge conjugates in the interval $1<\pt<12~{\rm GeV}/c$. The $\Xi^+_{\rm c}$ baryon was reconstructed via the decay channel $\Xi^{-}\pi^{+}\pi^{+}$, BR $=$ ($2.86 \pm 1.21 \pm 0.38$)\%~\cite{Li:2019atu}, together with its charge conjugate, in the interval $4<\pt<12~{\rm GeV}/c$.
The ratio $\rm {\rm BR}(\Xi_c^0\rightarrow \Xi^-e^+\nu_e)/\rm {\rm BR}(\Xi_c^0\rightarrow \Xi^{-}\pi^+)$ was also measured.
In the following, the notation $\Xi_{\rm c}$ is used to refer to both $\Xi^{0}_{\rm c}$ and $\Xi^{+}_{\rm c}$ states, if not differently specified.

A description of the ALICE detector and its performance are reported in Refs.~\cite{Alice_performance,Aamodt:2008zz}. The data used for these analyses were recorded with a minimum-bias trigger, based on coincident signals in the two scintillator arrays (V0) located on both sides of the interaction vertex. Offline selections, based on the V0 and Silicon Pixel Detector signals~\cite{ALICE:2021mgk}, were applied to remove background from beam--gas collisions. Pile-up events (less than 1\%~\cite{aliceLumi13TeVrun2}) containing multiple primary vertices were rejected. Only events with a reconstructed primary vertex position within $\pm$10~cm in the longitudinal direction from the nominal centre of the detector were used. With these requirements, $1.9 \times 10^9$ pp events were selected, corresponding to an integrated luminosity of $\mathcal{L}_{\rm int}=$~32.08 $\pm$ 0.51~${\rm nb^{-1}}$~\cite{aliceLumi13TeVrun2}.

Charged-particle tracks and particle-decay vertices were reconstructed in the central barrel using the Inner Tracking System (ITS) and the Time Projection Chamber (TPC), which are located inside a solenoidal magnet of field strength 0.5~T. The hadron (electron) selection criteria are the same as those reported in Ref.~\cite{ALICE:2021mgk} (\cite{Acharya:2017lwf}).
Particle identification (PID)
was performed using the information on the energy loss (d$E$/d$x$) through the TPC gas, and with the flight-time measurement of the Time-Of-Flight detector (TOF)~\cite{Adam:2016ilk}.
The $\Xi^{-}$ baryons were reconstructed from the decay chain $\Xi^{-} \rightarrow \pi^{-}\Lambda$, BR~$=$~($99.887 \pm 0.035$)\%, followed by $\Lambda \rightarrow {\rm\pi^{-} p}$, BR $=$ ($63.9 \pm 0.5$)\%~\cite{Zyla:2020zbs}. 
The $\Xi^{-}$ and $\Lambda$ baryons were reconstructed by exploiting their characteristic decay topologies as reported in Refs.~\cite{Acharya:2019kyh,Acharya:2017lwf}.

For the measurements in the hadronic decay channels, pions were selected according to the criteria described in Ref.~\cite{Acharya:2017jgo}. The $\rm \Xi_c$ candidates were reconstructed combining one or two pions, with the correct electric charge, to the selected $\Xi$ baryon.  
A Kalman-Filter vertexing algorithm~\cite{KalmanFilter} was used for the reconstruction of the $\rm \Xi_c^0\rightarrow \Xi^-\pi^+$ decay channel.
The package allows us to set constraints on the mass and on the production point of the reconstructed particles, using also information about the errors of daughter particle trajectories improving reconstruction accuracy of the mother particle. The mass constraint improves the mass and momentum reconstruction of the particle, while the production point constraint helps to determine whether the particle is coming from a certain vertex. These constraints were applied to each vertex and particle ($\Lambda$ and $\Xi$) in the decay chain reconstruction.
In the case of the $\rm \Xi_c^{+}$ baryon, the mean-proper lifetime $c\tau=132~\mu$m~\cite{Zyla:2020zbs} was exploited. The $\rm \Xi_c^{+}$ secondary vertex was reconstructed using only two pions having the same-sign charge, because the reconstructed $\rm \Xi$ trajectory has a much worse resolution when propagated to the primary vertex. Selections on the cosine of the pointing angle of the $\rm \Xi_{\rm c}^+$ to the primary vertex, the distance of closest approach between the two decay pions, and the decay length of the reconstructed secondary vertex were applied.
For the $\rm \Xi_{\rm c}^0$-baryon analysis, a multivariate technique based on the adaptive Boosted Decision Tree (BDT) algorithm in the Toolkit for Multivariate Data Analysis (TMVA)~\cite{Hocker:2007ht} was used.
The BDT algorithm was trained using reconstructed signal candidates obtained by simulating pp collisions with PYTHIA 8.2~\cite{Sjostrand:2014zea} and propagating the generated particles through the detector using the GEANT3 transport code~\cite{Brun:1994aa}, including a realistic description of the detector response and alignment during the data taking period. The background candidates were taken from data by selecting candidates with invariant mass in the intervals $2.17 < M < 2.39$ GeV/$c^2$ and $2.55 < M < 2.77$ GeV/$c^2$. 
The model was trained independently for each \pt interval with input variables related to the $\Xi^-$ decay topology and to the PID information of the decay tracks.
The $\Xi_{\rm c}$ raw yields were obtained from fits to the candidate invariant-mass distributions.
The signal peak was modelled with a Gaussian and the background was described by a linear function.

The $\rm \Xi_c^0\rightarrow \Xi^-e^+\nu_e$ analysis was performed using the technique reported in Ref.~\cite{Acharya:2017lwf}.
The $\Xi^0_{\rm c}$ candidates were defined from opposite-sign charge ${\rm e}\Xi$ pairs with an opening angle smaller than 90\degree. In order to reject electrons from photon conversions
occurring in the detector material, the electron-candidate tracks are required to have associated hits in the two innermost layers of the ITS~\cite{Acharya:2018upq,Acharya:2019mom}. Further rejection of background electrons originating from Dalitz decays of neutral mesons and photon conversions was performed using a technique based on the invariant mass of $\rm e^+e^-$ pairs~\cite{Acharya:2019lkh,Acharya:2019lkw}.
The electron (positron) candidates were paired with opposite-sign charge tracks
from the same event and are rejected if they form at least one $\rm e^{+}e^{-}$ pair with an invariant mass smaller than 50 MeV/$c^2$.
A correction for the misidentification probability was implemented, estimated to be 2\% by applying the algorithm to same-sign charge ${\rm e^{\pm}e^{\pm}}$ pairs.
The background in the ${\rm e^{+}\Xi^-}$ pair distribution is estimated by exploiting the fact that $\Xi^0_{\rm c}$ baryons decay into ${\rm e}^+\Xi^-\bar{\nu}_{\rm e}$, but not into ${\rm e}^-\Xi^-\bar{\nu}_{\rm e}$, while most of the background sources contribute equally to both samples. The yield of same-sign charge pairs is therefore used to estimate the background.
The $\Xi^0_{\rm c}$ raw yield was then obtained by subtracting the distribution of same-sign charge ${\rm e}\Xi$-pairs from the distribution of opposite-sign charge pairs, and integrating the invariant-mass distribution for $M({\rm e}\Xi)<2.5~{\rm GeV}/c^{2}$. 
The procedure was verified with PYTHIA 8.2~\cite{Sjostrand:2014zea} simulations and the GEANT 3 transport code. A similar procedure was adopted by the ARGUS and CLEO Collaborations~\cite{ARGUS:1992jnv, CLEO:1994aud}.
The same-sign charge pairs also contain a contribution from $\Xi^{0,-}_{\rm b}\rightarrow {\rm e}^-\Xi^-\bar{\nu}_{\rm e}$X decays not present in the distribution of opposite-sign charge pairs, leading to an oversubtraction. It was corrected for based on the assumptions reported in Ref.~\cite{Acharya:2017lwf} and ranges from 1\% to 4\%, depending on \pt.
The \pt distribution of $\rm e^{+}\Xi^{-}$ pairs was corrected for the missing momentum of the undetected neutrino using the Bayesian unfolding technique~\cite{DAgostini:1994fjx} implemented in the RooUnfold package~\cite{Adye:2011gm}. Additional information on the unfolding procedure is reported in the additional material~\cite{addmaterial}.

The raw yields were divided by the acceptance-times-efficiency for prompt hadrons, $({\rm Acc} \times \epsilon)_{\rm prompt}$, and were corrected for the beauty feed-down contribution. The $({\rm Acc} \times \epsilon)_{\rm prompt}$ corrections were obtained from a Monte Carlo simulation with the same configuration of the one used for the BDT training. The simulated $\Xi_{\rm c}$ \pt distributions were modified by a two step iterative procedure in order to mimic data. In the first step, the  $\Xi_{\rm c}$ reconstruction efficiency is obtained with the \pt distribution generated with PYTHIA~8.2. This $({\rm Acc} \times \epsilon)_{\rm prompt}$ is then used to calculate a first estimate of the $\Xi^0_{\rm c}$ \pt-differential spectrum. This first estimate is used to reweight the simulated $\Xi_{\rm c}$ \pt distributions, which is then used for the final computation of the $({\rm Acc} \times \epsilon)_{\rm prompt}$.
The $({\rm Acc}\times\epsilon)_{\rm prompt}$ increases with \pt from 0.6\% to 12\% depending on the particle and decay channel.
The contribution from beauty feed-down to the measured $\rm \Xi_c$ yields was subtracted.
The cross section of feed-down $\Xi_{\rm c}$ is calculated from the one of $\Lambda^+_{\rm c}$ originating from $\Lambda_{\rm b}^{0}$ decays (as described in Ref.~\cite{Acharya:2020uqi}) and scaled by the fraction of $\Xi_{\rm b}$ decaying in a final state with a $\Xi_{\rm c}$, which is taken to be about 50\% from the PYTHIA~8.2 generator~\cite{Sjostrand:2014zea}, and by the ratio of the measured \pt-differential yields of inclusive $\Xi_{\rm c}$ and prompt $\Lambda^+_{\rm c}$ baryons.
This procedure relies on the assumptions that the \pt shape of the cross sections of feed-down $\Lambda^+_{\rm c}$ and $\Xi_{\rm c}$ are similar, and that the ratio $\Xi_{\rm c}/\Lambda^+_{\rm c}$ is the same for inclusive and feed-down baryons.
The prompt fraction ($f_{\rm prompt}$) decreases with increasing \pt and it ranges from 0.99 at low \pt to 0.93 at high \pt.
To obtain the prompt $\Xi_{\rm c}$ cross sections, the corrected yields were divided by a factor of 2 to obtain the particle-antiparticle averaged yields, by the BR, by the widths of the \pt and $y$ intervals considered, and by $\mathcal{L}_{\rm int}$, as shown in Eq.~\ref{eq::corrected_yield}.

\begin{equation}
	\frac{\text{d}^2\sigma^{\Xi^0_{\rm c}}}{\text{d}p_\mathrm{T}\text{d}y}=\frac{1}{\rm BR}\times\frac{1}{2\Delta y\Delta p_\mathrm{T}}\times\frac{f_{\rm prompt} \times N^{\Xi^{0}_{\rm c}+\overline{\Xi}^{0}_{\rm c}}_\text{raw}}{({\rm Acc} \times\varepsilon)_{\rm prompt}}\times\frac{1}{\mathcal{L}_{\rm int}}.
	\label{eq::corrected_yield}
\end{equation}

Systematic uncertainties were estimated considering several sources.
For the hadronic decay channels, the systematic uncertainty on the raw-yield extraction was evaluated by repeating the fit of the invariant-mass distribution with varied fit interval, functional form of the background contribution, and width of the Gaussian function used to describe the signal peak. 
For the $\Xi^{0}_{\rm c}$ in the semileptonic decay channel, the raw-yield extraction systematic uncertainty was estimated by varying the selection criteria on the opening angle and on the invariant mass of the pair. The systematic uncertainties were defined as the RMS of the distribution of the signal yields obtained from these variations. The relative uncertainty on raw-yield extraction ranges from 7\% to 11\% depending on the \pt.
The uncertainty on the track reconstruction efficiency was evaluated by varying the track-selection criteria and by comparing the probability to prolong the tracks from the TPC to the ITS hits in data and simulations. A 5\% (7\%) uncertainty was assigned for the $\Xi^{0}_{\rm c}$ ($\Xi^{+}_{\rm c}$). 
The uncertainty on the selection efficiency originates mainly from imperfections in the description of the detector response and alignment in the simulation. It was estimated from the ratios of the corrected yields obtained by varying the BDT and topological selections applied; an uncertainty ranging from 2\% to 5\% was assigned.
The systematic uncertainty due to the shape of the $\Xi_{\rm c}$ \pt distributions used for the calculation of $({\rm Acc}\times\epsilon)_{\rm prompt}$ was estimated by considering different \pt shapes in the simulation, obtained by varying the weights mentioned above within their uncertainty~\cite{Acharya:2017lwf} and it amounts to 1\% for $\pt<3~{\rm GeV}/c$. 
The systematic uncertainty on the subtraction of feed-down from beauty-hadron decays was evaluated as in Ref.~\cite{Acharya:2020uqi} and additionally by scaling up the $\Xi_{\rm c}/\Lambda^+_{\rm c}$ ratio by a conservative factor of two and scaling it down to the $\Xi_{\rm b}^{\rm -}/\Lambda_{\rm b}^{\rm 0}$ ratio measured by the LHCb Collaboration~\cite{Aaij:2019ezy}, important in the case that BR($\Xi_{\rm b}^{0} \rightarrow \Xi_{\rm c}^{-}$X) is the same as BR($\Lambda_{\rm b}^{0} \rightarrow \Lambda_{\rm c}^{+}$X). The assigned
uncertainty ranges from 1\% to 9\% depending on $\pt$.
Additional uncertainties related only to the $\Xi^{0}_{\rm c}$ semileptonic decay channel were estimated as follows. 
The uncertainties related to the unfolding procedure were estimated by varying the number of iterations of the algorithm, the \pt range and the widths of the \pt intervals used in the Bayesian unfolding procedure, and the unfolding method itself to the Singular Value Decomposition~\cite{Hocker:1995kb}, and ranges from 2\% to 12\% depending on \pt.
The systematic uncertainty related to the oversubtraction due to the $\Xi_{\rm b}$ contribution in the same-sign charge ${\rm e}\Xi$ pairs was estimated by scaling the assumed $\Xi_{\rm b}$ momentum distribution by a conservative 50\%~\cite{Chatrchyan:2012xg}. A maximum of 2\% uncertainty was assigned at high $\pt$. 
A 2\% uncertainty was assigned to account for possible differences in the acceptance of $\rm e^{+}\Xi^{-}$ pairs in data and simulation, which is evaluated by performing the measurement in different rapidity intervals between $|y|<~0.5$ and 0.8.
The cross sections have an additional global normalisation uncertainty due to the uncertainties on the integrated luminosity~\cite{aliceLumi13TeVrun2} and the BRs~\cite{Zyla:2020zbs,Li:2019atu}.

\begin{figure}[!ht]
\centering
\includegraphics[width=0.45\textwidth]{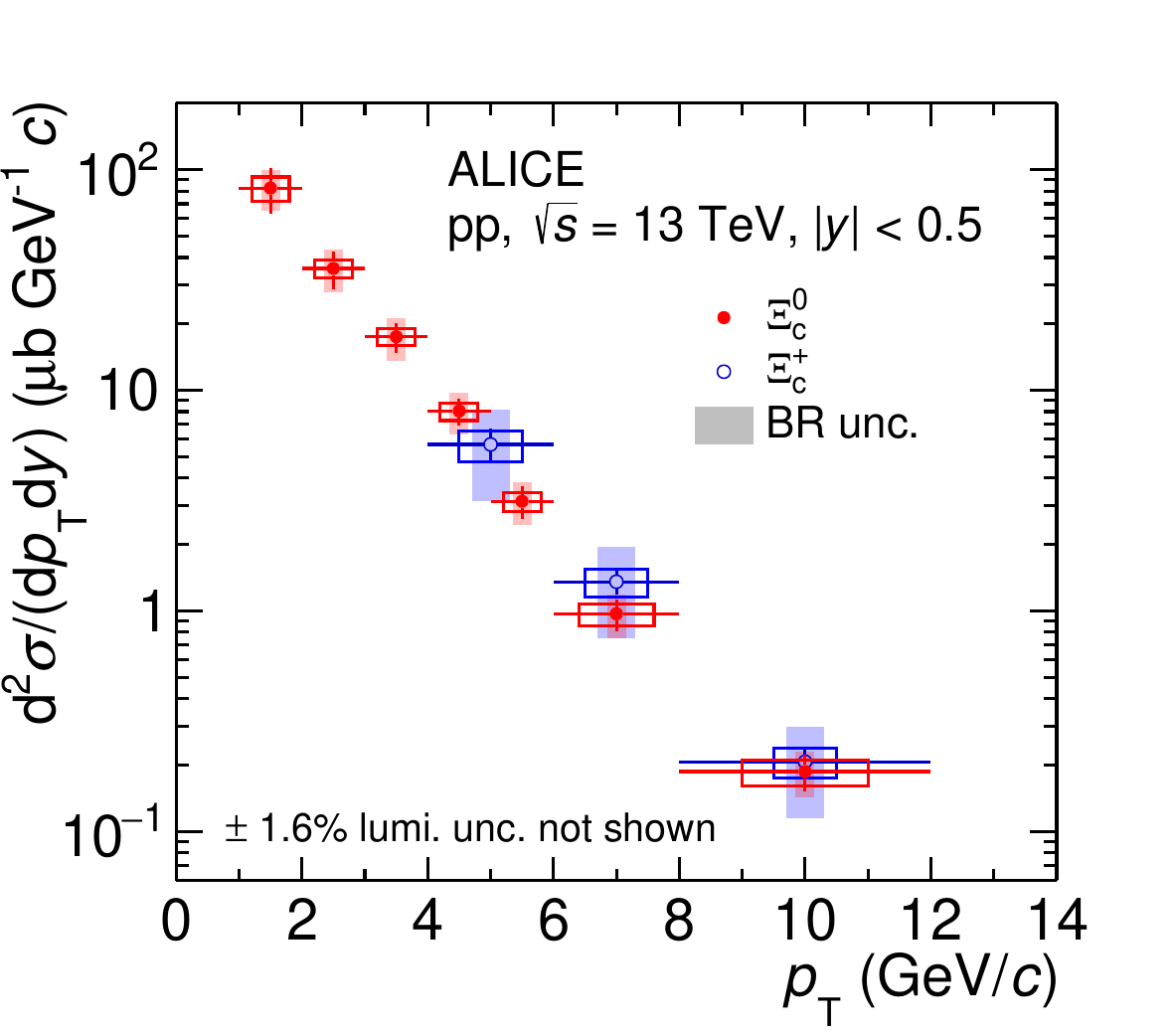}
\caption{Cross sections of prompt $\rm \Xi_c^0$ (full circles) and $\rm \Xi_c^+$ (open circles) baryons as a function of $p_{\rm T}$ in pp collisions at $\sqrt{s}=13$~TeV. The error bars and empty boxes represent the statistical and systematic uncertainties, respectively. The systematic uncertainties on the BR are shown as shaded boxes. 
}
\label{fig:CrossSection}
\end{figure}

The $\Xi^0_{\rm c}$ measurements in the two decay channels agree within statistical and uncorrelated systematic uncertainties~\cite{addmaterial}.
The results from the two decay channels were combined to obtain a more precise measurement of the prompt \pt-differential $\Xi^0_{\rm c}$-baryon cross section. The tracking and feed-down systematic uncertainties were propagated as correlated between the two measurements. 
Figure~\ref{fig:CrossSection} shows the average of the cross sections, computed considering as weights the inverse square of the relative statistical and \pt-uncorrelated systematic uncertainties~\cite{relativevar}. The prompt $\Xi^+_{\rm c}$-baryon cross section, also shown in Fig~\ref{fig:CrossSection}, is compatible within the uncertainties with the $\Xi^0_{\rm c}$ measurement.

\begin{figure}[ht!]
\begin{minipage}[c]{0.5\linewidth}
\centering
\includegraphics[width=.9\textwidth]{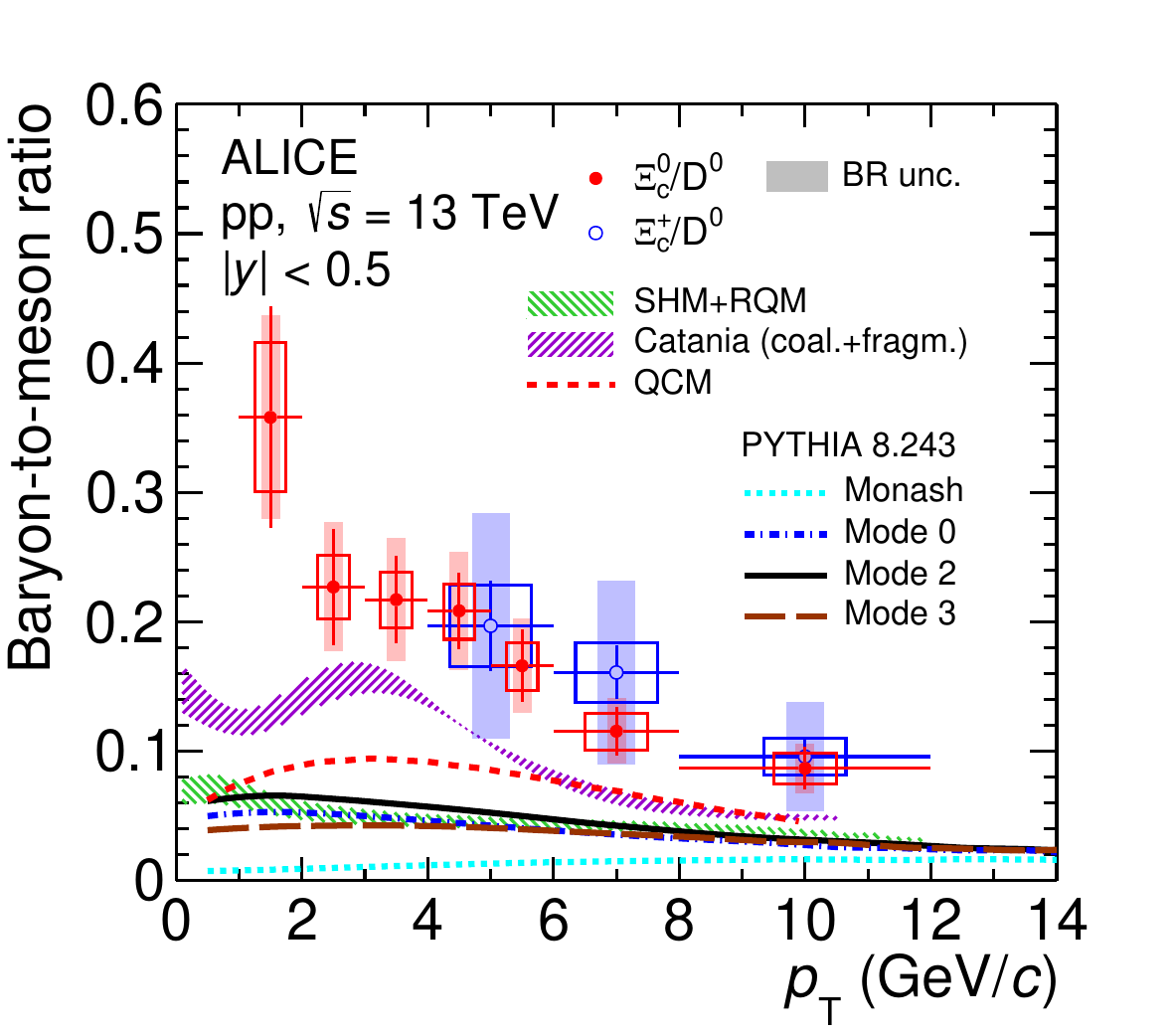}
\end{minipage}
\hspace{0.05cm}
\begin{minipage}[c]{0.5\linewidth}
\centering
\includegraphics[width=.9\textwidth]{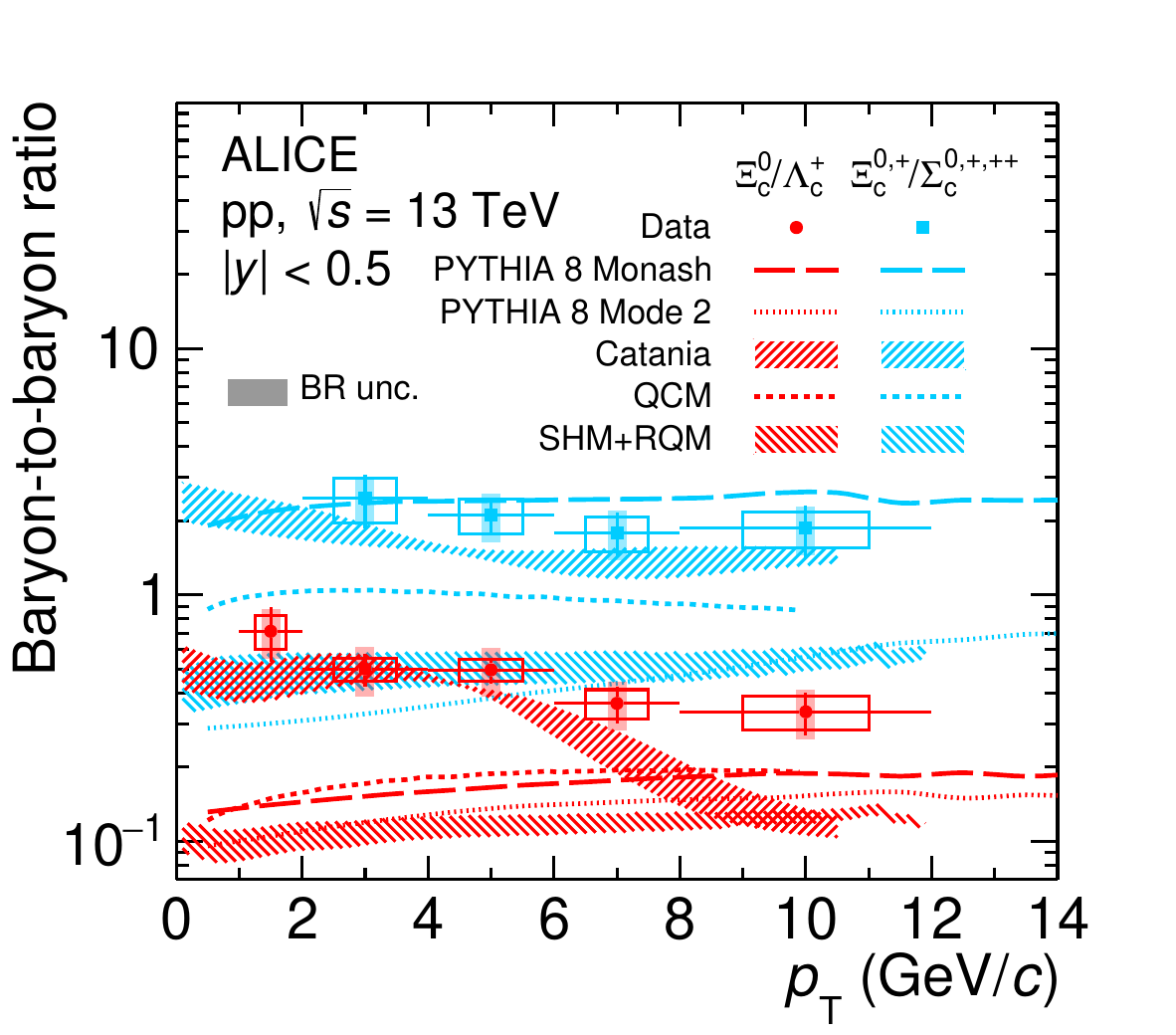}
\end{minipage}%
\caption{Left panel: $\rm \Xi_c^0/D^0$ and $\rm \Xi_c^+/D^0$ ratios as a function of $p_{\rm T}$ in pp collisions at $\sqrt{s}=13$~TeV.
Right panel: $\rm \Xi_c^0/\Lambda_c^+$ and $\rm \Xi_c^{0, +}/\Sigma_c^{0, +, ++}$ ratio as a function of $p_{\rm T}$.
The error bars and empty boxes represent the statistical and systematic uncertainties, respectively. The systematic uncertainties on the BR are shown as shaded boxes. The measurements are compared with model calculations (see text for detail).}
\label{fig2}
\end{figure}

The \pt-integrated cross sections in the measured \pt interval for the $\Xi_{\rm c}$ are $\text{d}\sigma^{\Xi^0_{\rm c}}_{\text{pp, 13 TeV}}/\text{d}y\big{|}^{(1 < \pt < 12 \text{~GeV}/c)}_{|y| < 0.5} = 149.2 \pm 20.7~\text{(stat)} \pm 35.5~\text{(syst)} \pm 2.4~\text{(lumi)}~\mu{\rm b}$ and $\text{d}\sigma^{\Xi^+_{\rm c}}_{\text{pp, 13 TeV}}/\text{d}y\big{|}^{(4 < \pt < 12 \text{~GeV}/c)}_{|y| < 0.5} = 14.9 \pm 2.0~\text{(stat)} \pm 6.6~\text{(syst)} \pm 0.2~\text{(lumi)}~\mu{\rm b}$. In calculating the \pt-integrated cross section and the ratio of the branching fractions, the systematic uncertainty related to unfolding, for the $\rm \Xi_c^0\rightarrow \Xi^-e^+\nu_e$, was considered as \pt uncorrelated and the other uncertainties as fully \pt correlated. For the hadronic decay channels, the uncertainty related to the raw-yield extraction was considered \pt uncorrelated, because the signal-over-background ratio is observed to largely vary as a function of \pt, while the others as fully \pt correlated.
The \pt-integrated $\Xi^0_{\rm c}$ cross section at midrapidity was obtained by extrapolating the visible cross section to the full \pt range. The \pt dependence of the Catania model~\cite{Minissale:2020bif}, which better describes the shape of the measured cross section with respect to other model calculations as seen in Fig.~\ref{fig2}, was used to calculate the extrapolation factor, which is $1.29^{+0.12}_{-0.08}$. The systematic uncertainty was estimated considering calculations~\cite{Christiansen:2015yqa,He:2019tik,Song:2018tpv} that describe the shape of the cross section in the measured \pt interval.
The \pt-extrapolated cross section for the $\Xi^0_{\rm c}$ is
$\text{d}\sigma^{\Xi^0_{\rm c}}_{\text{pp, 13 TeV}}/\text{d}y\big{|}_{|y| < 0.5} = 193.1~\pm~ 26.8~\text{(stat)}~\pm~46.0~\text{(syst)}~\pm~3.1~\text{(lumi)}$ $^{+17.6}_{-11.9}~\text{(extrap)}~\mu {\rm b}$. 

The measurement of the $\Xi^0_{\rm c}$-baryon cross sections, not corrected by the BRs, in the two different decay channels allowed the computation of the ${\rm BR}(\Xi^0_{\rm c} \rightarrow \Xi^{-}{\rm e}^{+}\nu_{\rm e})/ {\rm BR}(\Xi^0_{\rm c} \rightarrow \Xi^{-}{\pi}^{+})$ ratio.
The \pt-dependent ratio of the two measurements, which was observed to be flat in \pt~\cite{addmaterial}, was averaged over \pt using the inverse uncorrelated relative uncertainties as weights~\cite{relativevar}. 
The final systematic uncertainty on the ratio was obtained by summing in quadrature the $p_{\rm T}$-correlated and uncorrelated systematic uncertainties.
The measured ratio is ${\rm BR}(\Xi^0_{\rm c} \rightarrow \Xi^{-}{\rm e}^{+}\nu_{\rm e})/ {\rm BR}(\Xi^0_{\rm c} \rightarrow \Xi^{-}{\pi}^{+}) = 0.95 \pm 0.15~\text{(stat)} \pm 0.16~\text{(syst)}$~\cite{ALICE:2021bli_erratum}
and consistent with the one released by the Belle Collaboration~\cite{Belle:2021crz} within 1$\sigma$.

Figure~\ref{fig2} (left) shows the $\Xi_{\rm c}/{\rm D^0}$ ratios measured as a function of \pt. The systematic uncertainties related to the track-reconstruction efficiency, feed-down subtraction, and luminosity were propagated as correlated in the ratio.
The observed \pt dependence of the $\Xi_{\rm c}/{\rm D^0}$ ratio is similar to what was measured for the $\Lambda^{+}_{\rm c}/{\rm D^0}$ ratio~\cite{Acharya:2020uqi}, while the $\Xi_{\rm c}/{\rm D^0}$ ratio is generally lower. This result provides strong indications that the fragmentation functions of baryons and mesons differ significantly. 
The PYTHIA~8.2 event generator with the Monash tune~\cite{Sjostrand:2014zea}, and tunes that implement colour reconnection (CR) beyond the leading-colour approximation~\cite{Christiansen:2015yqa}, which lead to an increased baryon production, were compared to the measurements. 
The Monash tune significantly underestimates the data by a factor of 23--43 in the low-\pt region and by a factor of about 5 in the highest \pt interval.
All three CR modes give a similar magnitude and \pt-dependence of $\Xi_{\rm c}/{\rm D^0}$, and although they predict a larger baryon-to-meson ratio with respect to the Monash tune, they still underestimate the measured $\Xi_{\rm c}/{\rm D^0}$ ratio by a factor 4--9 for $\pt<4~{\rm GeV}/c$. 
The measured $\Xi_{\rm c}/{\rm D^0}$ ratio was also compared to a SHM~\cite{He:2019tik} that includes additional excited charm-baryon states not yet observed but predicted by the RQM~\cite{Ebert:2011kk} and by lattice QCD~\cite{Briceno:2012wt}. 
While this model describes the $\Lambda^+_{\rm c}/{\rm D^0}$ and $\Sigma^{0,+,++}_{\rm c}/{\rm D^0}$ ratios~\cite{Acharya:2020uqi,SigmacLambdac}, it underestimates the $\Xi_{\rm c}/{\rm D^0}$ ratio.
The measured ratios were also compared with models that include hadronisation via coalescence. In the quark (re-)combination mechanism (QCM)~\cite{Song:2018tpv}, the charm quark can pick up a comoving light antiquark or two comoving quarks to form a single-charm meson or baryon. The model does not describe the $\Xi_{\rm c}/{\rm D^0}$ ratio.
The Catania model~\cite{Minissale:2020bif,Plumari:2017ntm} implements charm-quark hadronisation via both coalescence and fragmentation, and it is the model that is closer to the measured ratio over the full \pt interval. 

The $\Xi_{\rm c}^0/\Lambda^+_{\rm c}$ and $\Xi^{0,+}_{\rm c}/\Sigma_{\rm c}^{0,+,++}$~\cite{SigmacLambdac} cross section ratios are reported in the right panel of Fig.~\ref{fig2}. The tracking, feed-down, and luminosity systematic uncertainties were propagated as correlated.
The $\Xi_{\rm c}^0/\Lambda^+_{\rm c}$ ratio is approximately 0.5 and within the current uncertainties there is no significant \pt dependence. All the PYTHIA~8.2 tunes, as well as the QCM, Catania, and the SHM+RQM models, do not describe the measured ratio.
To compute the $\Xi^{0,+}_{\rm c}/\Sigma_{\rm c}^{0,+,++}$, the $\rm \Xi_c^0$ was summed with the $\rm \Xi_c^+$ for $p_{\rm T}>4~{\rm GeV}/c$ and scaled by a factor of 2 in the interval $2<p_{\rm T}<4~{\rm GeV}/c$. The ratio is at approximately 2 and it is compatible with the Monash tune, which underestimates by a similar amount the $\Xi^{0,+}_{\rm c}$ and $\Sigma_{\rm c}^{0,+,++}$ cross sections~\cite{Acharya:2017lwf, SigmacLambdac}. The PYTHIA~8.2 tunes with CR and the SHM+RQM calculation also underestimate the measurement. The QCM model shows an almost flat value at unity, largely underestimating the measured ratio. The Catania model describes the data within the uncertainties.

In summary, measurements of the prompt charm-strange baryons $\Xi^+_{\rm c}$
and $\Xi^0_{\rm c}$ at midrapidity in pp collisions at $\sqrt{s}=$13~TeV were presented. The results 
pose important constraints to models of charm-quark hadronisation in pp collisions.
Finally, the ratio $\rm {\rm BR}(\Xi_c^0\rightarrow \Xi^-e^+\nu_e)/\rm {\rm BR}(\Xi_c^0\rightarrow \Xi^{-}\pi^+)$ was measured and consistent with the Belle's result within 1$\sigma$.


\newenvironment{acknowledgement}{\relax}{\relax}
\begin{acknowledgement}
\section*{Acknowledgements}

The ALICE Collaboration would like to thank all its engineers and technicians for their invaluable contributions to the construction of the experiment and the CERN accelerator teams for the outstanding performance of the LHC complex.
The ALICE Collaboration gratefully acknowledges the resources and support provided by all Grid centres and the Worldwide LHC Computing Grid (WLCG) collaboration.
The ALICE Collaboration acknowledges the following funding agencies for their support in building and running the ALICE detector:
A. I. Alikhanyan National Science Laboratory (Yerevan Physics Institute) Foundation (ANSL), State Committee of Science and World Federation of Scientists (WFS), Armenia;
Austrian Academy of Sciences, Austrian Science Fund (FWF): [M 2467-N36] and Nationalstiftung f\"{u}r Forschung, Technologie und Entwicklung, Austria;
Ministry of Communications and High Technologies, National Nuclear Research Center, Azerbaijan;
Conselho Nacional de Desenvolvimento Cient\'{\i}fico e Tecnol\'{o}gico (CNPq), Financiadora de Estudos e Projetos (Finep), Funda\c{c}\~{a}o de Amparo \`{a} Pesquisa do Estado de S\~{a}o Paulo (FAPESP) and Universidade Federal do Rio Grande do Sul (UFRGS), Brazil;
Ministry of Education of China (MOEC) , Ministry of Science \& Technology of China (MSTC) and National Natural Science Foundation of China (NSFC), China;
Ministry of Science and Education and Croatian Science Foundation, Croatia;
Centro de Aplicaciones Tecnol\'{o}gicas y Desarrollo Nuclear (CEADEN), Cubaenerg\'{\i}a, Cuba;
Ministry of Education, Youth and Sports of the Czech Republic, Czech Republic;
The Danish Council for Independent Research | Natural Sciences, the VILLUM FONDEN and Danish National Research Foundation (DNRF), Denmark;
Helsinki Institute of Physics (HIP), Finland;
Commissariat \`{a} l'Energie Atomique (CEA) and Institut National de Physique Nucl\'{e}aire et de Physique des Particules (IN2P3) and Centre National de la Recherche Scientifique (CNRS), France;
Bundesministerium f\"{u}r Bildung und Forschung (BMBF) and GSI Helmholtzzentrum f\"{u}r Schwerionenforschung GmbH, Germany;
General Secretariat for Research and Technology, Ministry of Education, Research and Religions, Greece;
National Research, Development and Innovation Office, Hungary;
Department of Atomic Energy Government of India (DAE), Department of Science and Technology, Government of India (DST), University Grants Commission, Government of India (UGC) and Council of Scientific and Industrial Research (CSIR), India;
Indonesian Institute of Science, Indonesia;
Istituto Nazionale di Fisica Nucleare (INFN), Italy;
Institute for Innovative Science and Technology , Nagasaki Institute of Applied Science (IIST), Japanese Ministry of Education, Culture, Sports, Science and Technology (MEXT) and Japan Society for the Promotion of Science (JSPS) KAKENHI, Japan;
Consejo Nacional de Ciencia (CONACYT) y Tecnolog\'{i}a, through Fondo de Cooperaci\'{o}n Internacional en Ciencia y Tecnolog\'{i}a (FONCICYT) and Direcci\'{o}n General de Asuntos del Personal Academico (DGAPA), Mexico;
Nederlandse Organisatie voor Wetenschappelijk Onderzoek (NWO), Netherlands;
The Research Council of Norway, Norway;
Commission on Science and Technology for Sustainable Development in the South (COMSATS), Pakistan;
Pontificia Universidad Cat\'{o}lica del Per\'{u}, Peru;
Ministry of Education and Science, National Science Centre and WUT ID-UB, Poland;
Korea Institute of Science and Technology Information and National Research Foundation of Korea (NRF), Republic of Korea;
Ministry of Education and Scientific Research, Institute of Atomic Physics and Ministry of Research and Innovation and Institute of Atomic Physics, Romania;
Joint Institute for Nuclear Research (JINR), Ministry of Education and Science of the Russian Federation, National Research Centre Kurchatov Institute, Russian Science Foundation and Russian Foundation for Basic Research, Russia;
Ministry of Education, Science, Research and Sport of the Slovak Republic, Slovakia;
National Research Foundation of South Africa, South Africa;
Swedish Research Council (VR) and Knut \& Alice Wallenberg Foundation (KAW), Sweden;
European Organization for Nuclear Research, Switzerland;
Suranaree University of Technology (SUT), National Science and Technology Development Agency (NSDTA) and Office of the Higher Education Commission under NRU project of Thailand, Thailand;
Turkish Energy, Nuclear and Mineral Research Agency (TENMAK), Turkey;
National Academy of  Sciences of Ukraine, Ukraine;
Science and Technology Facilities Council (STFC), United Kingdom;
National Science Foundation of the United States of America (NSF) and United States Department of Energy, Office of Nuclear Physics (DOE NP), United States of America.
\end{acknowledgement}

\bibliographystyle{utphys}   
\bibliography{bibliography}

\newpage
\appendix
\appendix

\section{Erratum to "Measurement of the Cross Sections of \Xiczero and \Xicplus Baryons and of the Branching-Fraction Ratio BR(\XicZeroToXiEleNu)/BR(\XicZeroToPiXi) in pp collisions at \thirteen" [Phys. Rev. Lett. 127, 272001 (2021)]}\label{erratum}


This erratum corrects the results reported in Ref.~\cite{ALICE:2021bli}, particularly the branching-fraction ratio between the two \Xiczero decay channels BR(\XicZeroToXiEleNu)/BR(\XicZeroToPiXi) due to an additional background contribution found in the analysis of the \Xiczero semileptonic decay channel (\XicZeroToXiEleNu).
Because the invariant mass of this decay channel cannot be fully reconstructed due to the presence of the neutrino in its decay chain, the \Xiczero candidates were estimated from opposite-sign charge \XiePair pairs and the background was evaluated from same-sign charge \XiePair pairs.
The raw yield was obtained by subtracting the same-sign charge distribution from the opposite-sign charge distribution within the invariant-mass range of $1.3 < M_{\XiePair} < 2.5$~\GeVmass. This approach assumes that the same-sign charge distribution accurately represents the total background of the opposite-sign charge distribution, leaving only the \XiePair pairs from the desired \XicZeroToXiEleNu decay after subtraction.
However, a publication from the Belle Collaboration~\cite{Belle:2021crz} reported additional decay channels producing opposite-sign charge \XiePair pairs: the channels
$ \Xiczero \to \mathrm{\Xi}^{\ast-}\eplus\nue \to \X\pizero\eplus\nue $ and
$ \Xicplus \to \mathrm{\Xi}^{\ast0}\eplus\nue \to \X\pip\eplus\nue $
(hereafter \textit{4-body decay} modes) generate opposite-sign charge \XiePair pairs with small mass difference to the \XicZeroToXiEleNu. They thus cannot be rejected by the conventional same-sign charge subtraction method.


\begin{figure}[b]
    \centering
    \includegraphics[width=0.65\textwidth]{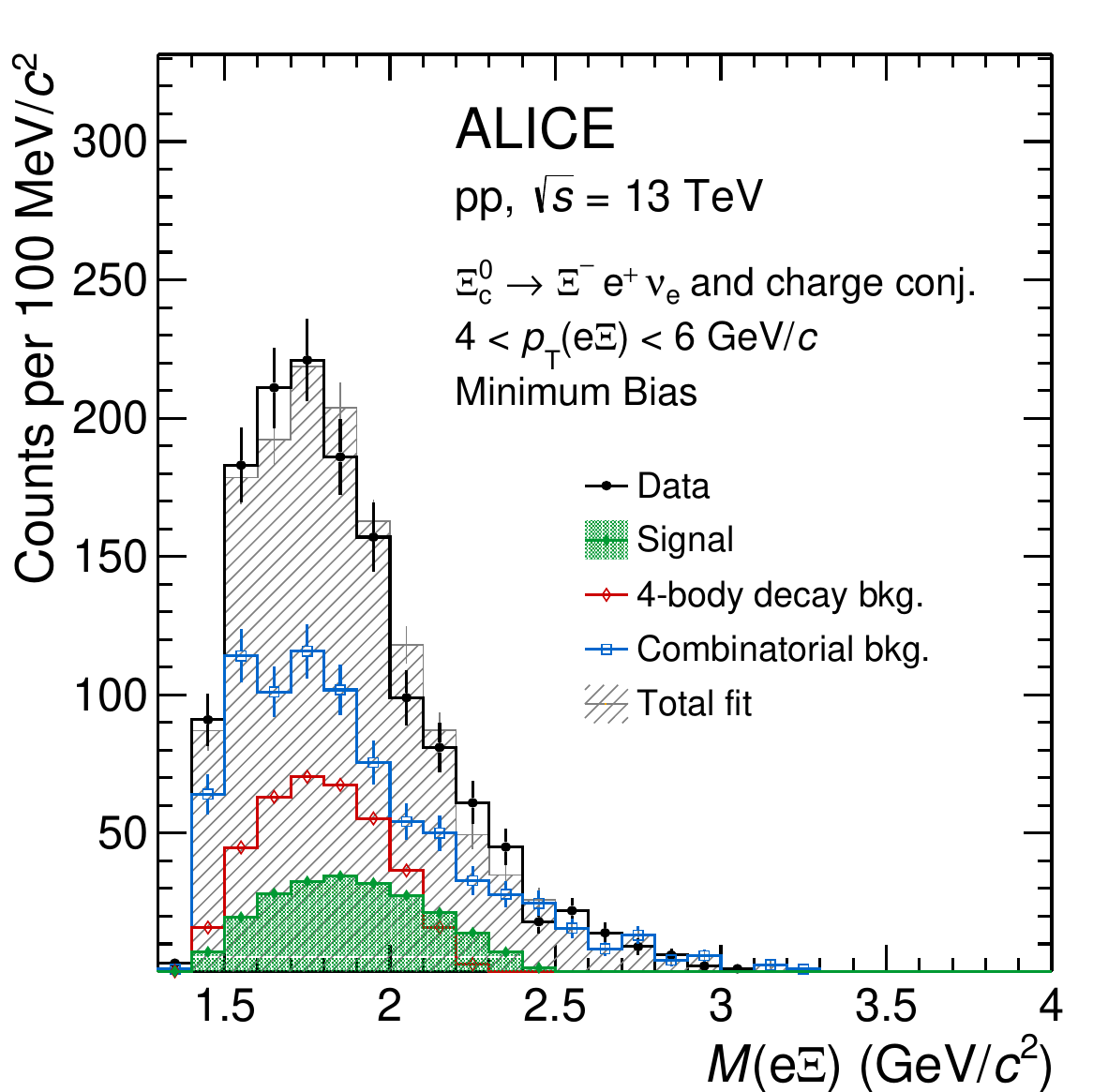}
    \caption{
    Template fit result for the interval of $4 < \pt < 6$ \GeVc: the black line indicates the $\Xi_{c}^{0}$ candidates, the green filled distribution indicates the $\Xi_{c}^{0} \to \Xi^{-} e^{+} \nu_{e}$ signal, the red line indicates the 4-body decay background, the blue line indicates the combinatorial background and the grey hashed line indicates the sum of each contribution, i.e., the total fit, respectively.}
    \label{fig:Invmass}
\end{figure}

Therefore, a template fit procedure was introduced to extract the raw yield considering the additional 4-body decay background contribution.
The template distributions considered were: i) signal from \XicZeroToXiEleNu obtained from Monte Carlo (MC) simulations, ii) background from 4-body decay modes obtained from the MC, and iii) background from random \XiePair combinations obtained from the invariant-mass distribution of same-sign charge \XiePair pairs of data. 
The MC simulation was based on the PYTHIA 8.243 event generator~\cite{Sjostrand:2014zea} and the GEANT 3 transport code~\cite{Brun:1994aa}, including a realistic description of the detector conditions during the data taking. During their generation, the \Xiczeroplus were forced to decay via semileptonic channels to enrich the MC sample.

The invariant-mass distributions of the signal and the 4-body decay background are very similar since the difference was only a soft pion from the resonance decay.
In addition, the limited sample of the \Xiczero candidates, in particular for \pt $>$ 6 \GeVc, did not allow us to perform the fit with the 4-body decay background normalisation as a free parameter. The fits were stabilized by adding a model constraint that the signal/(signal + 4-body decay background) ratio in \XiePair pairs follow the MC’s \pt dependence.
The \pt dependence appears due to additional pions from the 4-body decay mode, and it was found to be approximately linear in the seven \pt intervals of interest, in the range of \ptrange{1}{12} \GeVc based on PYTHIA. Therefore, the model was defined as follows:
\begin{equation}
    \alpha + \beta \times \pt
    \label{eq:frac34}
\end{equation}
where $\alpha$ was the signal/(signal + 4-body decay background) ratio at zero \pt and $\beta$ was the slope. During the fit procedure, $\alpha$ was set as a free parameter and $\beta$ was fixed from the PYTHIA simulation. The modelled distribution is:
\begin{equation}
    y = y_{0} \times
    (c_{\rm sig} \cdot y_{\rm sig} \,+\, c_{\rm 4\text{-}body} \cdot y_{\rm 4\text{-}body} \,+\, c_{\rm bg} \cdot y_{\rm bg})
    \label{eq:fit}
\end{equation}
where $y$ is the estimated yield of a mass bin in a \pt interval by the fit,
$y_{0}$ is the yield of the data,
$c_{\rm sig}$ is the fraction of the signal by the model, i.e., the Eq. \ref{eq:frac34},
$y_{\rm sig}$ is the yield of the signal obtained from the signal template,
$c_{\rm 4\text{-}body}$ is the fraction of the 4-body decay background by the model, i.e., (1 - $c_{\rm sig}$),
$y_{\rm 4\text{-}body}$ is the yield of the 4-body decay background,
$c_{\rm bg}$ is the fraction of the random background, and
$y_{\rm bg}$ is the yield of the random background, respectively.
During the fit process, the model in Eq. \ref{eq:fit} was introduced into the $c_{\rm sig}$ and the $c_{\rm 4\text{-}body}$ and works as a constraint to stabilise the fit across multiple \pt intervals of interest.

After obtaining the raw yield of \XiePair pairs from the desired \XicZeroToXiEleNu decay channel, the rest of the analysis procedure to acquire the branching ratio (BR) uncorrected cross section of \Xiczero, such as \pt unfolding, acceptance and efficiency \acceff correction, and non-prompt \Xiczero yield correction is the same as the one of the original paper~\cite{ALICE:2021bli}. 
Only the systematic uncertainty on the raw yield extraction is affected by this update. In the original analysis, the systematic uncertainty of the raw yield was measured by varying the upper limit of the invariant-mass distribution from 2.3 to 2.7 \GeVmass. In the updated analysis, the uncertainty was estimated via a MC closure test: multiple pseudo-data samples with a known (true) signal yield were created, and the template fit was performed on these pseudo-data samples to check the difference between the true signal yield and the measured yield.
To create a pseudo-data sample, each type of \XiePair pair was randomly sampled from its base template (e.g., signal template), with the same number of candidates as for data in each \pt interval, with the signal fraction varying from 5\% to 25\%.
Finally, the systematic uncertainty was estimated as the deviation between the true signal yield and the measured signal yield (1 - $N_{\rm measured}/{N_{\rm true}}$). The uncertainty of the model in Eq. \ref{eq:frac34} was also investigated by varying the slope parameter $\beta$. However, its effect was negligible.


The branching-fraction ratio between the two \Xiczero decay channels was recalculated using the updated result of the BR-uncorrected \Xiczero cross section from the \XicZeroToXiEleNu decay:
\begin{equation}
    \frac
    {\text{\rm BR\,(\XicZeroToXiEleNu)}}{\text{\rm BR\,(\XicZeroToPiXi)}}
    \,=\, 0.95 \, \pm \, 0.15 \; \text{(stat)} \, \pm \, 0.16 \; \text{(syst)}
    \label{eq:BRfrac}
\end{equation}
which is consistent with the Belle result of 0.73 $\pm$ 0.02 (stat) $\pm$ 0.04 (syst) within 1$\sigma$.

The BR corrected \Xiczero cross section reported in the original paper was a combination of results from \XicZeroToXiEleNu and \XicZeroToPiXi, using the inverse square of the relative statistical and uncorrelated systematic uncertainties as a weight. During the calculation, the contribution from the \XicZeroToXiEleNu was affected by a larger systematic uncertainty of the branching ratio of (1.8 $\pm$ 1.2)\%, in contrast to the (1.43 $\pm$ 0.32)\% for the \XicZeroToPiXi. Therefore, the impact of the 4-body decay background considered \XicZeroToXiEleNu cross section is negligible in the combined cross section result.

The branching-fraction ratio reported in the original paper~\cite{ALICE:2021bli} was used to normalise the measurement performed at \XicZeroToXiEleNu measurement in \pp collisions at\five~\cite{ALICE:2021psx}, by using the BR for the \XicZeroToXiEleNu decay obtained as 
\begin{equation}
    \text{\rm BR\,(\XicZeroToXiEleNu)} = 
    \text{\rm BR\,(\XicZeroToPiXi)}_{\rm PDG} \,\times\,
    \left[
    \frac{\text{\rm BR\,(\XicZeroToXiEleNu)}}{\text{\rm BR\,(\XicZeroToPiXi)}}
    \right]_{\rm ALICE}
    \label{eq:BRXic0SemiL}
\end{equation}
where the adopted branching-fraction ratio is the same as the one published in Ref.~\cite{ALICE:2021bli}, hence without subtraction of the 4-body decay background.
Since the raw yield in the corresponding paper was estimated using the same-sign charge \XiePair pairs subtraction method, the same contamination that caused a bias in the branching-fraction ratio is also present by the remaining 4-body decay background.
The \pt-differential production cross section was evaluated as
\begin{equation}
    \frac{\mathrm{d}^{2}\sigma}{\mathrm{d}{p_{\rm T}}\mathrm{d}y} =
    \frac{1}{2\Delta{p_{\rm T}}\Delta{p_{\rm y}}} \times
    \frac{N_{\rm raw}\,\cdot\,f_{\rm prompt}}{(\rm Acc\times\varepsilon)_{\rm prompt}} \times
    \frac{1}{\mathcal{L}_{\rm int}} \times
    \frac{1}{\rm BR}
    \label{eq:xSec}
\end{equation}
where $N_{\rm raw}$ is the raw yield obtained without the subtraction of the 4-body decay background, $f_{\rm prompt}$ and $\rm (Acc\times\varepsilon)_{\rm prompt}$ are the fraction of prompt \Xiczero baryons and the acceptance-times efficiency, respectively, $\mathcal{L}_{\rm int}$ is the integrated luminosity, while BR is the branching ratio for the \XicZeroToXiEleNu decay defined in Eq. \ref{eq:BRXic0SemiL}. Given that the 4-body decay background was not subtracted either in $N_{\rm raw}$ (numerator of Eq. \ref{eq:xSec}) or BR (denominator of Eq. \ref{eq:xSec}), and assuming that the relative contribution from the 4-body decay background is independent of the collision energy, the bias cancels out in the cross section calculation. Therefore, the impact is negligible.
%
%

\newpage
\section{The ALICE Collaboration}
\label{app:collab}
%
\begingroup
\small
\begin{flushleft}
S.~Acharya$^{\rm 143}$, 
D.~Adamov\'{a}$^{\rm 98}$, 
A.~Adler$^{\rm 76}$, 
J.~Adolfsson$^{\rm 83}$, 
G.~Aglieri Rinella$^{\rm 35}$, 
M.~Agnello$^{\rm 31}$, 
N.~Agrawal$^{\rm 55}$, 
Z.~Ahammed$^{\rm 143}$, 
S.~Ahmad$^{\rm 16}$, 
S.U.~Ahn$^{\rm 78}$
% ...

\section*{Affiliation notes}

$^{\rm I}$ Deceased\\
$^{\rm II}$ Also at: Italian National Agency for New Technologies, Energy and Sustainable Economic Development (ENEA), Bologna, Italy\\
$^{\rm III}$ Also at: Dipartimento DET del Politecnico di Torino, Turin, Italy\\
$^{\rm IV}$ Also at: M.V. Lomonosov Moscow State University, D.V. Skobeltsyn Institute of Nuclear, Physics, Moscow, Russia\\
$^{\rm V}$ Also at: Institute of Theoretical Physics, University of Wroclaw, Poland\\

\section*{Collaboration Institutes}

$^{1}$ A.I. Alikhanyan National Science Laboratory (Yerevan Physics Institute) Foundation, Yerevan, Armenia\\
$^{2}$ AGH University of Science and Technology, Cracow, Poland\\
$^{3}$ Bogolyubov Institute for Theoretical Physics, National Academy of Sciences of Ukraine, Kiev, Ukraine\\
$^{4}$ Bose Institute, Department of Physics  and Centre for Astroparticle Physics and Space Science (CAPSS), Kolkata, India\\
$^{5}$ Budker Institute for Nuclear Physics, Novosibirsk, Russia\\
$^{6}$ California Polytechnic State University, San Luis Obispo, California, United States\\
$^{7}$ Central China Normal University, Wuhan, China\\
$^{8}$ Centro de Aplicaciones Tecnol\'{o}gicas y Desarrollo Nuclear (CEADEN), Havana, Cuba\\
$^{9}$ Centro de Investigaci\'{o}n y de Estudios Avanzados (CINVESTAV), Mexico City and M\'{e}rida, Mexico\\
$^{10}$ Chicago State University, Chicago, Illinois, United States\\
$^{11}$ China Institute of Atomic Energy, Beijing, China\\
$^{12}$ Chungbuk National University, Cheongju, Republic of Korea\\
$^{13}$ Comenius University Bratislava, Faculty of Mathematics, Physics and Informatics, Bratislava, Slovakia\\
$^{14}$ COMSATS University Islamabad, Islamabad, Pakistan\\
$^{15}$ Creighton University, Omaha, Nebraska, United States\\
$^{16}$ Department of Physics, Aligarh Muslim University, Aligarh, India\\
$^{17}$ Department of Physics, Pusan National University, Pusan, Republic of Korea\\
$^{18}$ Department of Physics, Sejong University, Seoul, Republic of Korea\\
$^{19}$ Department of Physics, University of California, Berkeley, California, United States\\
$^{20}$ Department of Physics, University of Oslo, Oslo, Norway\\
$^{21}$ Department of Physics and Technology, University of Bergen, Bergen, Norway\\
$^{22}$ Dipartimento di Fisica dell'Universit\`{a} 'La Sapienza' and Sezione INFN, Rome, Italy\\
$^{23}$ Dipartimento di Fisica dell'Universit\`{a} and Sezione INFN, Cagliari, Italy\\
$^{24}$ Dipartimento di Fisica dell'Universit\`{a} and Sezione INFN, Trieste, Italy\\
$^{25}$ Dipartimento di Fisica dell'Universit\`{a} and Sezione INFN, Turin, Italy\\
$^{26}$ Dipartimento di Fisica e Astronomia dell'Universit\`{a} and Sezione INFN, Bologna, Italy\\
$^{27}$ Dipartimento di Fisica e Astronomia dell'Universit\`{a} and Sezione INFN, Catania, Italy\\
$^{28}$ Dipartimento di Fisica e Astronomia dell'Universit\`{a} and Sezione INFN, Padova, Italy\\
$^{29}$ Dipartimento di Fisica e Nucleare e Teorica, Universit\`{a} di Pavia, Pavia, Italy\\
$^{30}$ Dipartimento di Fisica `E.R.~Caianiello' dell'Universit\`{a} and Gruppo Collegato INFN, Salerno, Italy\\
$^{31}$ Dipartimento DISAT del Politecnico and Sezione INFN, Turin, Italy\\
$^{32}$ Dipartimento di Scienze e Innovazione Tecnologica dell'Universit\`{a} del Piemonte Orientale and INFN Sezione di Torino, Alessandria, Italy\\
$^{33}$ Dipartimento di Scienze MIFT, Universit\`{a} di Messina, Messina, Italy\\
$^{34}$ Dipartimento Interateneo di Fisica `M.~Merlin' and Sezione INFN, Bari, Italy\\
$^{35}$ European Organization for Nuclear Research (CERN), Geneva, Switzerland\\
$^{36}$ Faculty of Electrical Engineering, Mechanical Engineering and Naval Architecture, University of Split, Split, Croatia\\
$^{37}$ Faculty of Engineering and Science, Western Norway University of Applied Sciences, Bergen, Norway\\
$^{38}$ Faculty of Nuclear Sciences and Physical Engineering, Czech Technical University in Prague, Prague, Czech Republic\\
$^{39}$ Faculty of Science, P.J.~\v{S}af\'{a}rik University, Ko\v{s}ice, Slovakia\\
$^{40}$ Frankfurt Institute for Advanced Studies, Johann Wolfgang Goethe-Universit\"{a}t Frankfurt, Frankfurt, Germany\\
$^{41}$ Fudan University, Shanghai, China\\
$^{42}$ Gangneung-Wonju National University, Gangneung, Republic of Korea\\
$^{43}$ Gauhati University, Department of Physics, Guwahati, India\\
$^{44}$ Helmholtz-Institut f\"{u}r Strahlen- und Kernphysik, Rheinische Friedrich-Wilhelms-Universit\"{a}t Bonn, Bonn, Germany\\
$^{45}$ Helsinki Institute of Physics (HIP), Helsinki, Finland\\
$^{46}$ High Energy Physics Group,  Universidad Aut\'{o}noma de Puebla, Puebla, Mexico\\
$^{47}$ Hiroshima University, Hiroshima, Japan\\
$^{48}$ Hochschule Worms, Zentrum  f\"{u}r Technologietransfer und Telekommunikation (ZTT), Worms, Germany\\
$^{49}$ Horia Hulubei National Institute of Physics and Nuclear Engineering, Bucharest, Romania\\
$^{50}$ Indian Institute of Technology Bombay (IIT), Mumbai, India\\
$^{51}$ Indian Institute of Technology Indore, Indore, India\\
$^{52}$ Indonesian Institute of Sciences, Jakarta, Indonesia\\
$^{53}$ INFN, Laboratori Nazionali di Frascati, Frascati, Italy\\
$^{54}$ INFN, Sezione di Bari, Bari, Italy\\
$^{55}$ INFN, Sezione di Bologna, Bologna, Italy\\
$^{56}$ INFN, Sezione di Cagliari, Cagliari, Italy\\
$^{57}$ INFN, Sezione di Catania, Catania, Italy\\
$^{58}$ INFN, Sezione di Padova, Padova, Italy\\
$^{59}$ INFN, Sezione di Pavia, Pavia, Italy\\
$^{60}$ INFN, Sezione di Roma, Rome, Italy\\
$^{61}$ INFN, Sezione di Torino, Turin, Italy\\
$^{62}$ INFN, Sezione di Trieste, Trieste, Italy\\
$^{63}$ Inha University, Incheon, Republic of Korea\\
$^{64}$ Institute for Gravitational and Subatomic Physics (GRASP), Utrecht University/Nikhef, Utrecht, Netherlands\\
$^{65}$ Institute for Nuclear Research, Academy of Sciences, Moscow, Russia\\
$^{66}$ Institute of Experimental Physics, Slovak Academy of Sciences, Ko\v{s}ice, Slovakia\\
$^{67}$ Institute of Physics, Homi Bhabha National Institute, Bhubaneswar, India\\
$^{68}$ Institute of Physics of the Czech Academy of Sciences, Prague, Czech Republic\\
$^{69}$ Institute of Space Science (ISS), Bucharest, Romania\\
$^{70}$ Institut f\"{u}r Kernphysik, Johann Wolfgang Goethe-Universit\"{a}t Frankfurt, Frankfurt, Germany\\
$^{71}$ Instituto de Ciencias Nucleares, Universidad Nacional Aut\'{o}noma de M\'{e}xico, Mexico City, Mexico\\
$^{72}$ Instituto de F\'{i}sica, Universidade Federal do Rio Grande do Sul (UFRGS), Porto Alegre, Brazil\\
$^{73}$ Instituto de F\'{\i}sica, Universidad Nacional Aut\'{o}noma de M\'{e}xico, Mexico City, Mexico\\
$^{74}$ iThemba LABS, National Research Foundation, Somerset West, South Africa\\
$^{75}$ Jeonbuk National University, Jeonju, Republic of Korea\\
$^{76}$ Johann-Wolfgang-Goethe Universit\"{a}t Frankfurt Institut f\"{u}r Informatik, Fachbereich Informatik und Mathematik, Frankfurt, Germany\\
$^{77}$ Joint Institute for Nuclear Research (JINR), Dubna, Russia\\
$^{78}$ Korea Institute of Science and Technology Information, Daejeon, Republic of Korea\\
$^{79}$ KTO Karatay University, Konya, Turkey\\
$^{80}$ Laboratoire de Physique des 2 Infinis, Ir\`{e}ne Joliot-Curie, Orsay, France\\
$^{81}$ Laboratoire de Physique Subatomique et de Cosmologie, Universit\'{e} Grenoble-Alpes, CNRS-IN2P3, Grenoble, France\\
$^{82}$ Lawrence Berkeley National Laboratory, Berkeley, California, United States\\
$^{83}$ Lund University Department of Physics, Division of Particle Physics, Lund, Sweden\\
$^{84}$ Moscow Institute for Physics and Technology, Moscow, Russia\\
$^{85}$ Nagasaki Institute of Applied Science, Nagasaki, Japan\\
$^{86}$ Nara Women{'}s University (NWU), Nara, Japan\\
$^{87}$ National and Kapodistrian University of Athens, School of Science, Department of Physics , Athens, Greece\\
$^{88}$ National Centre for Nuclear Research, Warsaw, Poland\\
$^{89}$ National Institute of Science Education and Research, Homi Bhabha National Institute, Jatni, India\\
$^{90}$ National Nuclear Research Center, Baku, Azerbaijan\\
$^{91}$ National Research Centre Kurchatov Institute, Moscow, Russia\\
$^{92}$ Niels Bohr Institute, University of Copenhagen, Copenhagen, Denmark\\
$^{93}$ Nikhef, National institute for subatomic physics, Amsterdam, Netherlands\\
$^{94}$ NRC Kurchatov Institute IHEP, Protvino, Russia\\
$^{95}$ NRC \guillemotleft Kurchatov\guillemotright  Institute - ITEP, Moscow, Russia\\
$^{96}$ NRNU Moscow Engineering Physics Institute, Moscow, Russia\\
$^{97}$ Nuclear Physics Group, STFC Daresbury Laboratory, Daresbury, United Kingdom\\
$^{98}$ Nuclear Physics Institute of the Czech Academy of Sciences, \v{R}e\v{z} u Prahy, Czech Republic\\
$^{99}$ Oak Ridge National Laboratory, Oak Ridge, Tennessee, United States\\
$^{100}$ Ohio State University, Columbus, Ohio, United States\\
$^{101}$ Petersburg Nuclear Physics Institute, Gatchina, Russia\\
$^{102}$ Physics department, Faculty of science, University of Zagreb, Zagreb, Croatia\\
$^{103}$ Physics Department, Panjab University, Chandigarh, India\\
$^{104}$ Physics Department, University of Jammu, Jammu, India\\
$^{105}$ Physics Department, University of Rajasthan, Jaipur, India\\
$^{106}$ Physikalisches Institut, Eberhard-Karls-Universit\"{a}t T\"{u}bingen, T\"{u}bingen, Germany\\
$^{107}$ Physikalisches Institut, Ruprecht-Karls-Universit\"{a}t Heidelberg, Heidelberg, Germany\\
$^{108}$ Physik Department, Technische Universit\"{a}t M\"{u}nchen, Munich, Germany\\
$^{109}$ Politecnico di Bari and Sezione INFN, Bari, Italy\\
$^{110}$ Research Division and ExtreMe Matter Institute EMMI, GSI Helmholtzzentrum f\"ur Schwerionenforschung GmbH, Darmstadt, Germany\\
$^{111}$ Russian Federal Nuclear Center (VNIIEF), Sarov, Russia\\
$^{112}$ Saha Institute of Nuclear Physics, Homi Bhabha National Institute, Kolkata, India\\
$^{113}$ School of Physics and Astronomy, University of Birmingham, Birmingham, United Kingdom\\
$^{114}$ Secci\'{o}n F\'{\i}sica, Departamento de Ciencias, Pontificia Universidad Cat\'{o}lica del Per\'{u}, Lima, Peru\\
$^{115}$ St. Petersburg State University, St. Petersburg, Russia\\
$^{116}$ Stefan Meyer Institut f\"{u}r Subatomare Physik (SMI), Vienna, Austria\\
$^{117}$ SUBATECH, IMT Atlantique, Universit\'{e} de Nantes, CNRS-IN2P3, Nantes, France\\
$^{118}$ Suranaree University of Technology, Nakhon Ratchasima, Thailand\\
$^{119}$ Technical University of Ko\v{s}ice, Ko\v{s}ice, Slovakia\\
$^{120}$ The Henryk Niewodniczanski Institute of Nuclear Physics, Polish Academy of Sciences, Cracow, Poland\\
$^{121}$ The University of Texas at Austin, Austin, Texas, United States\\
$^{122}$ Universidad Aut\'{o}noma de Sinaloa, Culiac\'{a}n, Mexico\\
$^{123}$ Universidade de S\~{a}o Paulo (USP), S\~{a}o Paulo, Brazil\\
$^{124}$ Universidade Estadual de Campinas (UNICAMP), Campinas, Brazil\\
$^{125}$ Universidade Federal do ABC, Santo Andre, Brazil\\
$^{126}$ University of Cape Town, Cape Town, South Africa\\
$^{127}$ University of Houston, Houston, Texas, United States\\
$^{128}$ University of Jyv\"{a}skyl\"{a}, Jyv\"{a}skyl\"{a}, Finland\\
$^{129}$ University of Kansas, Lawrence, Kansas, United States\\
$^{130}$ University of Liverpool, Liverpool, United Kingdom\\
$^{131}$ University of Science and Technology of China, Hefei, China\\
$^{132}$ University of South-Eastern Norway, Tonsberg, Norway\\
$^{133}$ University of Tennessee, Knoxville, Tennessee, United States\\
$^{134}$ University of the Witwatersrand, Johannesburg, South Africa\\
$^{135}$ University of Tokyo, Tokyo, Japan\\
$^{136}$ University of Tsukuba, Tsukuba, Japan\\
$^{137}$ Universit\'{e} Clermont Auvergne, CNRS/IN2P3, LPC, Clermont-Ferrand, France\\
$^{138}$ Universit\'{e} de Lyon, CNRS/IN2P3, Institut de Physique des 2 Infinis de Lyon , Lyon, France\\
$^{139}$ Universit\'{e} de Strasbourg, CNRS, IPHC UMR 7178, F-67000 Strasbourg, France, Strasbourg, France\\
$^{140}$ Universit\'{e} Paris-Saclay Centre d'Etudes de Saclay (CEA), IRFU, D\'{e}partment de Physique Nucl\'{e}aire (DPhN), Saclay, France\\
$^{141}$ Universit\`{a} degli Studi di Foggia, Foggia, Italy\\
$^{142}$ Universit\`{a} di Brescia, Brescia, Italy\\
$^{143}$ Variable Energy Cyclotron Centre, Homi Bhabha National Institute, Kolkata, India\\
$^{144}$ Warsaw University of Technology, Warsaw, Poland\\
$^{145}$ Wayne State University, Detroit, Michigan, United States\\
$^{146}$ Westf\"{a}lische Wilhelms-Universit\"{a}t M\"{u}nster, Institut f\"{u}r Kernphysik, M\"{u}nster, Germany\\
$^{147}$ Wigner Research Centre for Physics, Budapest, Hungary\\
$^{148}$ Yale University, New Haven, Connecticut, United States\\
$^{149}$ Yonsei University, Seoul, Republic of Korea\\

\bigskip 

\end{flushleft} 
\endgroup  
\end{document}